\documentclass[aps,prl,twocolumn,footinbib,groupedaddress,amsmath,superscriptaddress]{revtex4}
\usepackage{amsmath}
\usepackage{graphicx}
\usepackage{amsfonts}
\usepackage{amssymb}
\usepackage{hyperref}
\usepackage{color}
\usepackage{mathrsfs}
\usepackage{isomath}
\usepackage{amsmath}
\usepackage{amsthm}
\usepackage{epstopdf}
\usepackage{txfonts}

\begin{document}

\title{Multipartite Gaussian steering: monogamy constraints and quantum cryptography applications}
\author{Yu Xiang}
\email{xiangy.phy@pku.edu.cn}
\address{State Key Laboratory of Mesoscopic Physics, School of Physics, Peking University, Collaborative Innovation Center of Quantum Matter, Beijing 100871, China}
\address{Collaborative Innovation Center of Extreme Optics, Shanxi University, Taiyuan, Shanxi 030006, China}
\author{Ioannis Kogias}
\email{john$_$k$_$423@yahoo.gr}
\address{School of Mathematical Sciences, University of Nottingham, Nottingham NG7 2RD, United Kingdom}
\author{Gerardo Adesso}
\email{gerardo.adesso@nottingham.ac.uk}
\address{School of Mathematical Sciences, University of Nottingham, Nottingham NG7 2RD, United Kingdom}
\author{Qiongyi He}
\email{qiongyihe@pku.edu.cn}
\address{State Key Laboratory of Mesoscopic Physics, School of Physics, Peking University, Collaborative Innovation Center of Quantum Matter, Beijing 100871, China}
\address{Collaborative Innovation Center of Extreme Optics, Shanxi University, Taiyuan, Shanxi 030006, China}


\begin{abstract}
We derive laws for the distribution of quantum steering among different parties in multipartite Gaussian states under Gaussian measurements. We prove that a monogamy relation akin to the generalized Coffman-Kundu-Wootters inequality holds quantitatively for a recently introduced measure of Gaussian steering. We then define the residual Gaussian steering, stemming from the monogamy inequality, as an indicator of collective steering-type correlations. For pure three-mode Gaussian states, the residual  acts a quantifier of genuine multipartite steering, and is interpreted operationally in terms of the guaranteed key rate in the task of secure quantum secret sharing. Optimal resource states for the latter protocol are identified, and their possible experimental implementation discussed. Our results pin down the role of multipartite steering for quantum communication.
\end{abstract}

\date{\today}

\maketitle
With the imminent debacle of Moore's law, and the constant need for faster and more reliable processing of information, quantum technologies are set to radically change the landscape of modern communication and computation. A successful and secure quantum network relies on quantum correlations distributed and shared over many sites
\cite{qinternet}. Different kinds of multipartite quantum correlations have been considered as valuable resources for various applications in quantum communication tasks. Multipartite entanglement \cite{vanLoockgen,branetwork,RevModPhys.81.865,Entanglementdetection,adesso2006CVent,3mpra,qubitmultient} and multipartite Bell nonlocality \cite{PhysRevD.35.3066,PhysRevLett.88.170405,PhysRevLett.89.060401,GerardoGenuineBell} are two well known instances and have received extensive attention in recent developments of quantum information theory, as well as  in other branches of modern physics. There has been substantial experimental progress in engineering and detection of both such correlations, by using e.g.~photons \cite{shalm2013threephoton,8photonGuo,eight-photonent,threephoton-Bell,erven2014experimental}, ions \cite{14-Qubit}, or continuous variable (CV) systems \cite{Seiji2012programmable,shanxitrip,shanxiprl,tokyo10000}. However, as an intermediate type of quantum correlation between entanglement and Bell nonlocality, multipartite quantum steering \cite{PhysRevLett.115.010402,armstrong2015multipartite} still defies a complete understanding. In consideration of the intrinsic relevance of the notion of steering to the foundational core of quantum mechanics, it has become a worthwhile objective to deeply explore the characteristics of multipartite steering distributed over many parties, and to establish what usefulness to multiuser quantum communication protocols can such a resource provide, where bare entanglement is not enough and Bell nonlocality may not be accessible.

The concept of quantum steering was originally introduced by Schr\"{o}dinger  \cite{Schrodinger}  to describe the ``spooky action-at-a-distance" effect noted in the Einstein-Podolsky-Rosen (EPR) paradox \cite{Einstein,Reid89,rmpReid}, whereby local measurements performed on one party apparently adjust (steer) the state of another distant party. Recently identified as a distinct type of nonlocality \cite{Wiseman07,Wiseman07pra},
quantum steering is thus a directional form of quantum correlations, characterized by its inherent asymmetry between the parties \cite{PhysRevLett.106.130402,handchen2012observation,One-way,PhysRevA.92.032107,QYClassifying,Quantification,Hierarchy}.
Additionally, steering allows verification of entanglement, without assumptions of the full trust of reliability of equipment at all of the nodes of a communication network \cite{PhysRevA.80.032112}. Steering is then a natural resource for one-sided device-independent quantum key distribution  \cite{1sdI-QKD,walk}. For bipartite systems, a comprehensive quantitative investigation of quantum steering has been recently proposed \cite{PhysRevLett.112.180404,Kogias:15,costa2015quantification,QYtele} and tested in several systems \cite{saunders2010experimental,PhysRevX.2.031003,1367-2630-14-5-053030,sun2015experimental,PhysRevLett.113.140402,prydenew,NoGauss1}. Comparatively little is known about steering in multipartite scenarios. For instance, Refs.~\cite{QYGenuine,Ryan,cavalcanti-nc} derived criteria to detect genuine multipartite steering, and Ref.~\cite{reidmonogamy} presented some limitations on joint quantum steering in tripartite systems.

In this Rapid Communication we focus on steerability of multipartite Gaussian states of CV systems by Gaussian measurements, a physical scenario which closely aligns with the traditional EPR paradox, and which is of primary relevance for experimental implementations \cite{ourreview,weedbrookrmp,cvbook}. In order to investigate the shareability of Gaussian steering from a quantitative perspective \cite{Quantification}, we establish {\it monogamy} relations imposing constraints on the degree of bipartite EPR steering that can be shared among $N$-mode CV systems in pure Gaussian states, in analogy with the Coffman-Kundu-Wootters (CKW) monogamy inequality for entanglement \cite{ckw,adesso2006CVent,3mpra,hiroshima,strongckw,GerardoRenyi,lancien_2016}.  We further propose an indicator of collective steering-type correlations, the \textit{residual Gaussian steering} (RGS), stemming from the laws of steering monogamy, that is shown to act as a quantifier of genuine multipartite steering for pure three-mode Gaussian states. Finally, we show how the RGS acquires an operational interpretation in the context of a partially device-independent quantum secret sharing (QSS) protocol \cite{Hillery,weedbrook,QSSus,notecleve,Cleve}. Specifically, taking into account arbitrary eavesdropping and potential cheating strategies of some of the parties \cite{QSSus}, the achievable key rate of the protocol is shown to admit tight lower and upper bounds which are simple linear functions of the RGS. This in turn allows us to characterize optimal resources for CV QSS in terms of their multipartite steering.

\paragraph{Monogamy of Gaussian steering.} A fundamental property of entanglement, that has profound applications in quantum communication, is known as monogamy \cite{ckw,barbaramono,lancien_2016}. Any two quantum systems that are maximally entangled with each other, cannot be entangled (or, even, classically correlated) with any other third system.
Therefore, entanglement cannot be freely shared among different parties.
In their seminal paper  \cite{ckw}, CKW derived a monogamy inequality that quantitatively describes this phenomenon for any finite entanglement shared among arbitrary three-qubit states $\rho$:
${\cal C}^2_{A:(BC)}\left( \rho \right) \geq {\cal C}^2_{A:B}\left( \rho \right)+{\cal C}^2_{A:C}\left( \rho \right)$,
where ${\cal C}^2_{A:(BC)}\left( \rho \right)$ is the squared concurrence, quantifying the amount of bipartite entanglement across the bipartition $A:(BC)$. Osborne and Verstraete later generalized the CKW monogamy inequality to $n$ qubits \cite{strongckw}.
For CV systems, however, both the quantification and the study of the distribution of entanglement constitute in general a considerably harder problem. Remarkably, if one focuses on the theoretically and practically relevant class of Gaussian states, various results similar to the qubit case have been derived, using different entanglement measures \cite{adesso2006CVent,3mpra,hiroshima,GerardoRenyi,strongckwcv,ourreview}. Of particular interest to us will be the fact that the Gaussian R\' enyi-2 entanglement monotone ${\cal E}_{A:B} \left(\rho_{AB} \right)$, which quantifies entanglement of bipartite Gaussian states $\rho_{AB}$, has been shown to obey a CKW-type monogamy inequality for all $m$-mode Gaussian states $\rho_{A_1 \ldots A_m}$ with covariance matrix (CM) $\sigma_{A_1 \ldots A_m}$ \cite{GerardoRenyi},
\begin{equation}
\label{renyi}
{\cal E}_{A_k : \left(A_1,\ldots,A_{k-1},A_{k+1},\dots,A_m\right)} \left(\sigma_{A_1 \ldots A_m}\right)  - \sum_{j \neq k} {\cal E}_{A_k : A_j}\left(\sigma_{A_1 \ldots A_m}\right) \geq 0,
\end{equation}
where each $A_j$ comprises one mode only. Recall that the $2m \times 2m$ CM $\sigma_{A_1  \ldots A_m}$ of a $m$-mode state $\rho_{A_1  \ldots A_m}$ has elements ${\sigma _{ij}} = {\rm tr} \big[ {{{\{ {{{\hat R}_i},{{\hat R}_j}} \}}_ + }\ {\rho}} \big]$, where
$\hat R = (\hat x_1,\hat p_1, \ldots ,\hat x_m, \hat p_m)^{\sf T}$ is the vector collecting position and momentum operators of each mode, satisfying canonical commutation relations $[{{{\hat R}_i},{{\hat R}_j}} ] = i{(\Omega_{A_1  \ldots A_m})_{ij}}$, with $(\Omega_{A_1 \ldots A_m})  = \omega^{\oplus m}$ and $\omega =  {{\ 0\ \ 1}\choose{-1\ 0}}$ being the single-mode symplectic form \cite{ourreview}.

Quantum steering is a type of correlation that allows for entanglement certification in a multi-mode bipartite state $\rho_{AB}$ even when one of the parties' devices, say Bob's, are completely uncharacterized (untrusted). In this case, we say that Bob can steer Alice's local state \cite{Wiseman07,Wiseman07pra}. Keeping our focus on Gaussian states and measurements \cite{Quantification}, the question, thus, naturally arises: is steering monogamous?
Intuitively one would expect that there should exist limitations on the distribution of steering-type correlations, since steering is only a stronger form of the already monogamous entanglement.
A first answer to this question was recently given by Reid \cite{reidmonogamy}, who showed that, under restrictions to measurements and detection criteria involving up to second order moments, if a single-mode party $A$ can be steered by a single-mode party $B$ then no other single-mode party $C$ can simultaneously steer $A$. This was recently generalized to the case of parties $B$ and $C$ comprising an arbitrary number of modes \cite{Kimono,asimon}. Ref.~\cite{reidmonogamy} also discussed other monogamy relations for steering and nonlocality both in discrete and CV systems.

In the following we provide general quantitative CKW-type limitations to the distribution of Gaussian steering among many parties. For our purposes, we will focus on a recently proposed Gaussian steering measure \cite{Quantification}, ${\cal G}^{B\to A}\left( \sigma_{AB} \right)$, which quantifies how much party $B$ can steer party $A$ in a Gaussian state with CM $\sigma_{AB}$ by Gaussian measurements. In particular,  we now show that the Gaussian steering measure $\cal G$ is monogamous, hence satisfies a CKW-type monogamy inequality in direct analogy with entanglement. Consider an arbitrary
(pure or mixed)
$m$-mode Gaussian state $\rho_{A_1 \ldots A_m}$
with CM $\sigma_{A_1 \ldots A_m}$,
where each party $A_j$ comprises a single mode $(n_j=1,\,\, \forall j=1,\ldots, m)$.
Then, the following inequalities hold, $\forall \,k=1,\ldots,m$:
\begin{eqnarray}
\label{monog1}
\!\!\!\!\!\!{\cal G}^{(A_1,\ldots,A_{k-1},A_{k+1},\dots,A_m) \to A_k}(\sigma_{A_1 \ldots A_m}) -\! \sum_{j \neq k} {\cal G}^{A_j \to A_k}(\sigma_{A_1 \ldots A_m}) \!\!&\geq&\!\! 0,\\
\label{monog2}
\!\!\!\!\!\!{\cal G}^{A_k \to (A_1,\ldots,A_{k-1},A_{k+1},\dots,A_m)}(\sigma_{A_1 \ldots A_m}) -\! \sum_{j \neq k} {\cal G}^{A_k \to A_j}(\sigma_{A_1 \ldots A_m}) \!\!&\geq&\!\! 0.
\end{eqnarray}

\begin{figure}
\hspace{-1mm}\includegraphics[width=0.49\columnwidth]{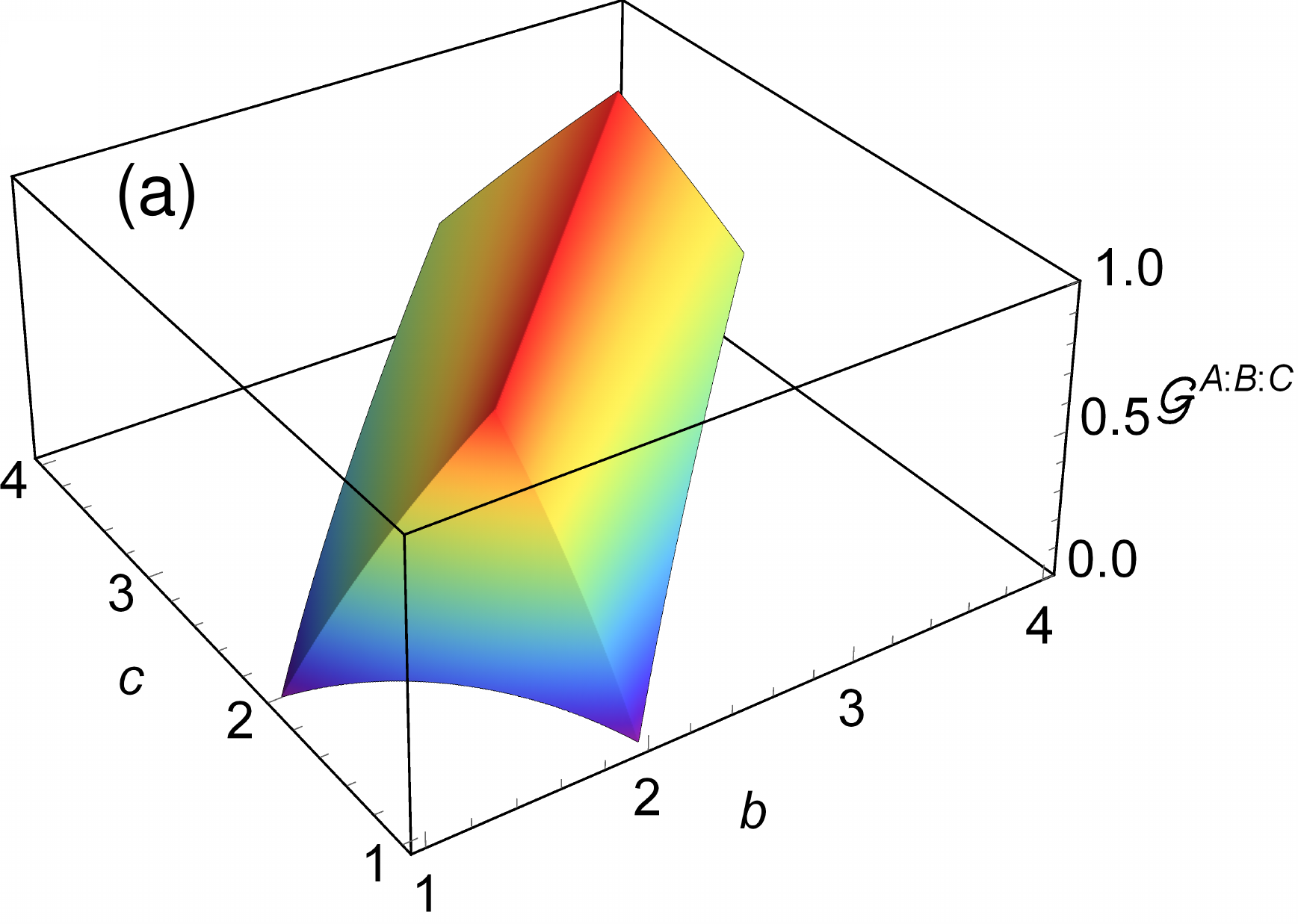}
\includegraphics[width=0.51\columnwidth]{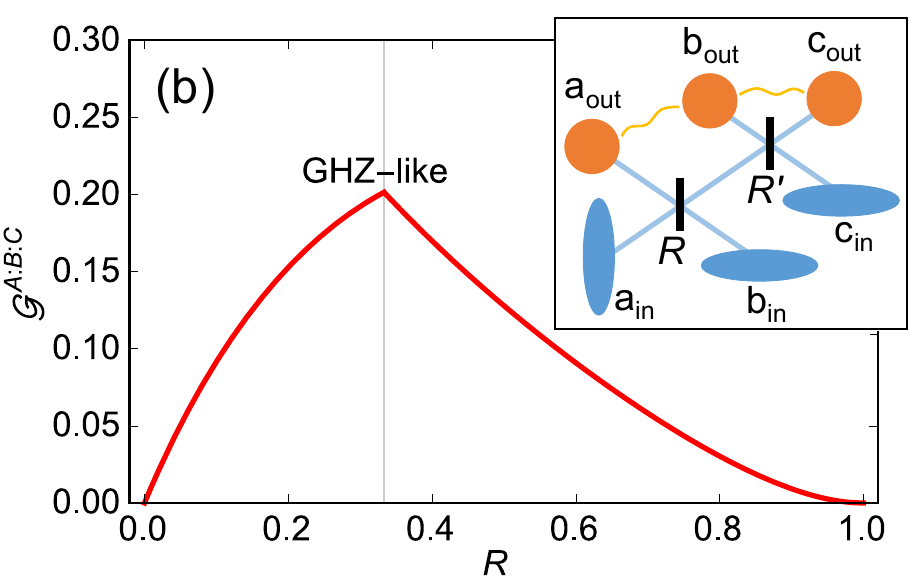}\hspace{-1mm}
\caption{Residual tripartite Gaussian steering ${\cal G}^{A:B:C}$ for pure three-mode Gaussian states with CM $\sigma^{\rm pure}_{ABC}$ (a) with fixed $a=2$ (local variance of subsystem $A$), and (b) generated  by three squeezed vacuum fields at $-3$ dB injected in two beamsplitters with reflectivities $R$ and $R'$ (see inset), setting $R'=1/2$ to obtain $b=c$; the permutationally invariant GHZ-like state ($a=b=c$) is obtained at $R=1/3$.}\label{Fig1}
\end{figure}

For pure states with CM $\sigma_{A_1 \ldots A_m}^{\rm pure}$, the proof is straightforward. 
Namely, recall from  \cite{Quantification} that the leftmost terms of \eqref{monog1}, \eqref{monog2} and \eqref{renyi} all coincide on pure states. On the other hand, for the marginal states of any two modes $i$ and $j$ one has ${\cal E}_{A_i : A_j}\left(\sigma_{A_1 \ldots A_m} ^{\rm pure}\right) \geq {\cal G}^{A_i \to A_j}\left( \sigma_{A_1 \ldots A_m}^{\rm pure}\right)$ \cite{Quantification}. Inequalities \eqref{monog1} and \eqref{monog2} then follow readily from the monogamy inequality \eqref{renyi} for Gaussian entanglement. The full proof of the above inequalities for general mixed states is deferred to the Appendix.

The monogamy relations just derived in this work impose fundamental restrictions to the distribution of Gaussian steering among multiple parties in fully quantitative terms.
To analyze these  in more detail, let us focus on a tripartite scenario, in which the monogamy inequalities take the simpler form,
\begin{align}\label{monogtrip1}
 &{\cal G}^{(AB) \rightarrow C} \left( \sigma_{ABC} \right) - {\cal G}^{A \rightarrow C}\left( \sigma_{ABC} \right) - {\cal G}^{B \rightarrow C}\left( \sigma_{ABC} \right) \geq 0, \\
 \label{monogtrip2}
 & {\cal G}^{C \rightarrow (AB)} \left( \sigma_{ABC} \right) - {\cal G}^{C \rightarrow A}\left( \sigma_{ABC} \right) - {\cal G}^{C \rightarrow B} \left( \sigma_{ABC} \right) \geq 0.
\end{align}
As in the original CKW inequality, these inequalities enjoy a very appealing interpretation: the degree of steering (by Gaussian measurements) exhibited by the state when all three parties are considered (i.e., ${\cal G}^{(AB) \rightarrow C}>0$, or, ${\cal G}^{C \rightarrow (AB)}>0)$ can be larger that the sum of the degrees of steering exhibited by the individual pairs. On a more extreme level, there exist quantum states where parties $A$ and $B$ cannot individually steer party $C$, i.e., $ {\cal G}^{A \rightarrow C} =  {\cal G}^{B \rightarrow C} = 0 $, but collectively they can, i.e., ${\cal G}^{(AB) \rightarrow C}>0$. We will see the importance of this type of correlations later when we discuss applications to QSS. We remark that the monogamy inequality (\ref{monogtrip1}) realizes a crucial nontrivial strengthening of Result 5 in \cite{reidmonogamy}, which can be recast as  ${\cal G}^{(AB) \rightarrow C} \left( \sigma_{ABC} \right) - \frac12 {\cal G}^{A \rightarrow C}\left( \sigma_{ABC} \right) - \frac12 {\cal G}^{B \rightarrow C}\left( \sigma_{ABC} \right) \geq 0$  in our notation. On the other hand, the reverse monogamy relation (\ref{monog2}) settles an open question raised in the same work \cite{reidmonogamy}.

The residuals of the subtractions in \eqref{monogtrip1}, \eqref{monogtrip2} quantify steering-type correlations that correspond to a collective property of the three parties, not reducible to the properties of the individual pairs. We proceed by investigating this quantitatively in a mode-invariant way. In analogy with what done for entanglement \cite{adesso2006CVent,3mpra,GerardoRenyi}, we can calculate the residuals from the monogamy inequalities \eqref{monogtrip1} or \eqref{monogtrip2} and minimise them over all mode permutations. It turns out that, in the paradigmatic case of pure three-mode Gaussian states with CM $\sigma_{ABC}^{\rm pure}$ ($m=3$), we obtain the same quantity (RGS) from either \eqref{monogtrip1} or \eqref{monogtrip2}, regardless of the steering direction. 
Explicitly, denoting by $\langle i,j,k\rangle$ any cyclic permutation of $A,B,C$, the RGS for three-mode pure Gaussian states with CM $\sigma_{ABC}^{\rm pure}$ is defined as
\begin{subequations}
\begin{align} \label{residual1}
 {\cal G}^{A:B:C}\left( \sigma_{ABC}^{\rm pure} \right) & = \min_{\langle i,j,k\rangle}
\left\{{{\cal G}^{(jk) \rightarrow i}  - {\cal G}^{j \rightarrow i} - {\cal G}^{k \rightarrow i}}\right\}\\
\label{residual2}
& = \min_{\langle i,j,k\rangle}
\left\{{{\cal G}^{i \rightarrow (jk)} - {\cal G}^{i \rightarrow j} - {\cal G}^{i \rightarrow k}}\right\}\\
& = \ln \left[ \min \left\{  b c/a,  ca/b,  ab/c  \right\} \right],
\end{align}
\end{subequations}
where $a=\sqrt{\det \sigma_A}$, $b=\sqrt{\det \sigma_B}$, and $c=\sqrt{\det \sigma_C}$ are local symplectic invariants (with $|b-c|+1 \leq a \leq b+c-1$),  fully determining the CM $\sigma_{ABC}^{\rm pure}$ in standard form \cite{3mpra,GerardoRenyi}.

The RGS ${\cal G}^{A:B:C}$ is a monotone under Gaussian local operations and classical communication, as one can prove analogously to the case of the residual entanglement of Gaussian states \cite{adesso2006CVent,3mpra,GerardoRenyi,Quantification,Kogias:15}. Furthermore, finding a non-zero value of the RGS  certifies genuine tripartite steering, as defined by He and Reid \cite{QYGenuine}, since a sufficient requirement to violate the corresponding biseparable model for pure states is the demonstration of steering in all directions: $\left(BC\right) \rightarrow A,\, \left(AC \right) \rightarrow B $ and $\left(AB\right) \rightarrow C$. We can then regard the RGS as a meaningful quantitative indicator of genuine tripartite steering for pure three-mode Gaussian states under Gaussian measurements.

In Fig.~\ref{Fig1}(a) we plot the RGS as a function of $b$ and $c$ for a given $a$. An elementary analysis reveals that the RGS  ${\cal G}^{A:B:C}$ is maximized on bisymmetric states with $b=c \geq a$, i.e., when the states are steerable across any global split of the three modes  and also  $B \leftrightarrow C$ steerable, but no other steering exists between any two parties. In this case, the genuine tripartite steering ${\cal G}^{A:B:C}$ reduces to the collective steering ${\cal G}^{(BC) \rightarrow A} = {\cal G}^{A \rightarrow (BC)} = \ln a$. This quantitative analysis completes the existing picture of quantum correlations in pure three-mode Gaussian states, together with the cases of tripartite Bell nonlocality in terms of maximum violation of the Svetlichny inequality \cite{GerardoGenuineBell} and genuine tripartite entanglement in terms of Gaussian R\' enyi-2 entanglement \cite{GerardoGenuineBell}. Bisymmetric states maximize all three forms of nonclassical correlations; compare e.g.~our Fig.~\ref{Fig1}(a) with Fig.~1(a)--(b) in \cite{GerardoGenuineBell}.

Figure~\ref{Fig1}(b) presents the RGS measure for Gaussian states generated by three squeezed vacuum fields (one in momentum, two in position) with experimentally feasible squeezing parameter $r=0.345$ (i.e., $3$ dB of squeezing) \cite{shanxitrip,8dB,10dB} injected at two beamsplitters with reflectivities $R$ and $R'$ as depicted in the inset of Fig.~\ref{Fig1}(b), setting $R'=1/2$ so that  $a=\sqrt{1+2R(1-R)(\cosh 4r-1)}$, $b=c=\sqrt{[1+R^2-(R^2-1) \cosh 4r]/2}$. When $R=1/3$, one can generate a permutationally invariant Greenberger-Horne-Zeilinger (GHZ)-like state with $a=b=c$ \cite{branetwork}. As one might expect, the latter states maximize the RGS in this case.


\paragraph{Operational connections to quantum secret sharing.}

Secret sharing \cite{shamir, Blakley} is a conventional cryptographic protocol in which  a dealer (Alice)  wants to share a secret with two players, Bob and Charlie, but with one condition: Bob and Charlie should be unable to individually access the secret (which may involve highly confidential information) and their collaboration would be required in order to prevent wrongdoings. 

QSS schemes \cite{Hillery, Karlsson, notecleve} have been proposed to securely accomplish this task, by exploiting multipartite entanglement to secure and split the classical secret among the players in a single go. Very recently, we provided an unconditional security proof for entanglement-based QSS protocols in a companion paper \cite{QSSus}.
In our scheme, the goal of the dealer is to establish a secret key with a joint degree of freedom of the players. The players can only retrieve Alice's key and decode the classical secret by collaborating and communicating to each other their local measurements to form the joint variable.  The unconditional security of these schemes stems from the utilized partially device-independent setting, treating the dealer as a trusted party with characterized devices, and the (potentially, dishonest) players as untrusted parties whose measuring devices are described as black boxes. Given this intrinsically asymmetric separation of roles, one would expect that multipartite steering be closely related to the security figure of merit of QSS. Here we prove such a connection quantitatively.

To start with, let us assume that the dealer, Alice, and the players, Bob and Charlie, all perform homodyne measurements of the quadratures $\hat{x}_i, \hat{p}_i$ with outcomes $X_i,P_i$, with $i=A,B,C$, on the shared tripartite state. Following \cite{QSSus}, a guaranteed (asymptotic) secret key rate for the QSS protocol (extracted from the correlations of Alice's momentum detection $P_A$ and a joint variable $\bar{P}$ for Bob and Charlie) to provide security against external eavesdropping is given by
$K^{A \rightarrow \{B,C\}}_{E} \geq - \ln\big(  {e \sqrt{ V_{P_A |\bar{P}} V_{X_A |\bar{X}}}} \big)$,
while the key rate providing {\it unconditional} security against both eavesdropping and dishonest actions of the players is
\begin{equation}\label{devetakfinalfinal}
\begin{split}
K^{A \rightarrow \{B,C\}}_{\rm{full}} \geq   -\ln \left( e \sqrt{ V_{P_A |\bar{P}} \cdot  \max \{ V_{X_A |X_C}, V_{X_A |X_B}   \} }   \right).
\end{split}
\end{equation}
Here, $V_{P_A |\bar{P}} = \int  d\bar{P}\, p(\bar{P}) \left( \langle P_A^2 \rangle_{\bar{P}} - \langle P_A \rangle^2_{\bar{P}} \right) $ is the minimum inference variance of Alice's momentum outcome given the players' joint outcome $\bar{P}$, and similarly for the other variances.
A tripartite shared state $\rho_{ABC}$ whose correlations result in nonzero values of the right-hand side of   (\ref{devetakfinalfinal}) can be regarded a useful resource for unconditionally secure QSS.

We focus on pure three-mode Gaussian states with CM $\sigma^{\rm pure}_{ABC}$ in standard form, fully specified by the local invariants $a,b,c$ as before. Our first observation is that  $K_E$ is directly quantified by the collective steering,  ${\cal G}^{(BC)\to A} \left(\sigma^{\rm pure}_{ABC}\right) = \max \left\{0, \frac{1}{2} \ln \frac{\det \sigma_{BC}}{\det \sigma_{ABC}} \right\}$. For the considered class of states, one has indeed $\frac{\det \sigma_{ABC}}{\det \sigma_{BC}}=4  V_{P_A |\bar{P}} V_{X_A |\bar{X}}=1/a^2$, where the joint variables were chosen to have the linear form $\bar{X}=g_X X_B + h_X X_C$ and  $\bar{P}=g_P P_B + h_P P_C$, with the real constants $g_{X(P)}, h_{X(P)}$ optimized as to minimize the inferred variances $V_{X_A |\bar{X}},V_{P_A |\bar{P}}$; see also \cite{walk,Quantification}. Putting everything together, we get: $
K_E^{A \to \{B,C\}} \left(\sigma^{\rm pure}_{ABC}\right)\geq \max\left\{0,\,{\cal G}^{(BC)\to A} \left(\sigma^{\rm pure}_{ABC}\right) - \ln\frac{e}{2}\right\}.
$

We can now define a mode-invariant QSS key rate bound $K_{\rm full}^{A:B:C}$ that takes into account eavesdropping and potential dishonesty of the players, by minimizing the right-hand side of Eq.~(\ref{devetakfinalfinal}) over the choice of the dealer, i.e., over permutations of $A$, $B$, and $C$. A nonzero value of the figure of merit $K_{\rm full}^{A:B:C}(\sigma_{ABC})$ on a tripartite Gaussian state with CM $\sigma_{ABC}$ guarantees the usefulness of the state for unconditionally secure QSS, for any possible assignment of the roles. For pure three-mode Gaussian states, the mode-invariant key rate $K_{\rm full}^{A:B:C}(\sigma^{\rm pure}_{ABC})$ can be evaluated explicitly (although its lengthy expression is omitted here) and analyzed in the physical space of the parameters  $a,b,c$. We find that $K_{\rm full}^{A:B:C}(\sigma^{\rm pure}_{ABC})$ admits {\it exact linear upper and lower bounds} as a function of the RGS ${\cal G}^{A:B:C}(\sigma^{\rm pure}_{ABC})$, for all states with standard form CM $\sigma^{\rm pure}_{ABC}$:
\begin{equation}\label{boundqss}
\frac{{\cal G}^{A:B:C}(\sigma^{\rm pure}_{ABC})}2 - \ln\frac{e}{2} \leq  K_{\rm full}^{A:B:C}(\sigma^{\rm pure}_{ABC}) \leq {\cal G}^{A:B:C}(\sigma^{\rm pure}_{ABC}) - \ln\frac{e}{2}\,.
\end{equation}

\begin{figure}
\includegraphics[width=0.7\columnwidth]{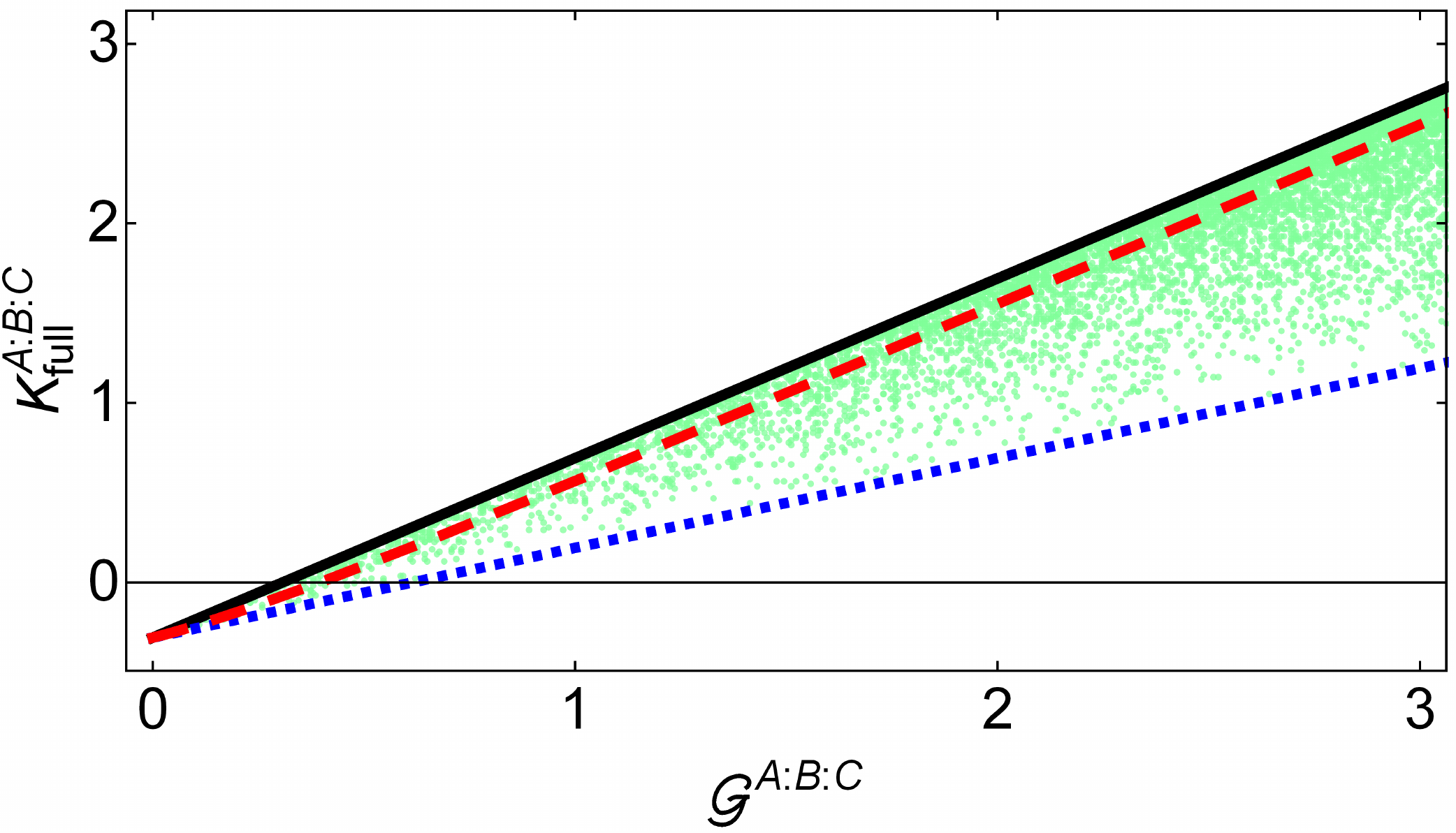}
\caption{Mode-invariant secure QSS key rate versus  RGS for $10^5$  pure three-mode Gaussian states (dots); see text for details on the lines.}\label{Fig2}
\end{figure}

The bounds are illustrated in Fig.~\ref{Fig2} together with a numerical exploration of $10^5$ randomly generated pure three-mode Gaussian states. Remarkably, the bounds are tight, and families of states saturating them can be readily provided.
Specifically, the lower (dotted blue) boundary is spanned by states with $a \geq 1$, $b=c=(a+1)/2$; 
conversely, the upper (solid black) boundary is spanned by states with $a \geq 1$, $b=c \rightarrow \infty$. 
While these cases are clearly extremal, GHZ-like states (dashed red), specified by $a=b=c$ and producible as discussed in Fig.~\ref{Fig1}(b),
nearly maximize the QSS key rate at fixed RGS, thus arising as convenient practical resources for the considered task, independently of the distribution of trust. Indeed, a squeezing level of $4.315$ dB, referring to the scheme of Fig.~\ref{Fig1}(b), is required to ensure a nonzero key rate using these states. This is well within the current experimental feasibility, since up to $10$ dB of squeezing has been demonstrated \cite{8dB,10dB}. In general, by imposing nonnegativity of the lower bound in (\ref{boundqss}), we find that  $K_{\rm full}^{A:B:C}(\sigma^{\rm pure}_{ABC})>0$ for all pure three-mode Gaussian states with RGS ${\cal G}^{A:B:C}(\sigma^{\rm pure}_{ABC})>2 \ln(e/2) \approx 0.614$. Our analysis reveals that partially device-independent QSS is empowered by multipartite steering, yielding a direct operational interpretation for the RGS in terms of the guaranteed key rate of the protocol.

\paragraph{Discussion and conclusion.}
We have proven that a recently proposed measure of quantum steering under Gaussian measurements \cite{Quantification,Kogias:15} obeys CKW-type monogamy inequalities for all  Gaussian states of any number of modes. We remark that monogamy extends in fact to arbitrary  non-Gaussian states under Gaussian measurements, as it is established solely at the level of covariance matrices. Notice however that resorting to non-Gaussian measurements can lead to extra steerability even for Gaussian states \cite{NoGauss1,NoGauss2}, and might allow circumventing some monogamy constraints \cite{reidmonogamy,Kimono,asimon}. 

In the important case of pure three-mode Gaussian states, we demonstrate that the residual steering emerging from the laws of monogamy can act as a quantifier of genuine tripartite steering. The latter measure is endowed with an operational interpretation, as it is shown to provide tight bounds on the mode-invariant key rate of a partially device-independent QSS protocol, whose unconditional security has been very recently investigated \cite{QSSus}. Our study, combined with \cite{QSSus}, provides practical recipes demonstrating that an implementation of QSS secure against eavesdropping and potentially dishonest players  is feasible with current technology using tripartite Gaussian states and Gaussian measurements \cite{notearm}.



\begin{acknowledgments}
{\it Note added.} After completion of this work, monogamy inequalities for multipartite Gaussian steering in the case of more than one mode per party have been investigated in \cite{LL}.

{\it Acknowledgments.} YX and QH acknowledge the support of the National Natural Science Foundation of China under Grants No.~11274025,  No.~61475006, and  No.~61675007. IK and GA acknowledge funding from the European Research Council under Starting Grant No.~637352 (GQCOP). GA thanks R.~Simon and L.~Lami for useful discussions.

\end{acknowledgments}

\paragraph{Appendix A: Proof of (\ref{monog1}).}
It suffices to prove the inequality for tripartite states as in (\ref{monogtrip1}),
 with $C$ being a single mode and $A$, $B$ comprising arbitrary number of modes. One can then apply iteratively this inequality to obtain the corresponding $m$-partite one (\ref{monog1}).

To do so, recall that from \cite{reidmonogamy,Kimono,asimon} it is impossible for $A$ and $B$ to simultaneously steer the one-mode party $C$, that is, ${\cal G}^{A \rightarrow C}\left( \sigma_{ABC} \right) > 0$ implies ${\cal G}^{B \rightarrow C}\left( \sigma_{ABC} \right) = 0$ (and vice versa). Therefore, the monogamy relation (\ref{monogtrip1}) reduces to ${\cal G}^{(AB) \rightarrow C} \left( \sigma_{ABC} \right) - {\cal G}^{A \rightarrow C}\left( \sigma_{ABC} \right) \geq 0$ (or the analogous expression with swapped $A \leftrightarrow B$), which holds true because the Gaussian steering measure (for one-mode steered party $C$) is nonincreasing under local Gaussian operations on the steering party $(AB)$ \cite{Kogias:15}, which include discarding  $B$ (or $A$). This proves Eq.~(\ref{monog1}) for any $m$-mode CM   $\sigma_{A_1 \ldots A_m}$. \hfill $\Box$


\paragraph{Appendix B: Proof of (\ref{monog2}).}
In this case we have to recall the explicit expression of the Gaussian steering measure
\cite{Kogias:15}, defined for a bipartite  ($n_A+n_B$)-mode state with CM $\sigma_{AB}$ as
\begin{equation*}\label{GSAtoB}
{\cal G}^{A \to B}(\sigma_{AB})=
\left\{
  \begin{array}{ll}
    0, &\hspace*{-2.2cm} \hbox{$\bar{\nu}^{AB\backslash A}_j \geq 1$ $\forall j=1,\ldots,n_B$\ ;} \\
    -\sum_{j:\bar{\nu}^{AB\backslash A}_j<1} \ln\left(\bar{\nu}^{AB\backslash A}_j\right), & \hbox{otherwise,}
  \end{array}
\right.
\end{equation*}
where $\big\{\bar{\nu}^{AB\backslash A}_j\big\}_{j=1}^{n_B}$ denote the symplectic eigenvalues of the Schur complement $\bar{\sigma}_{AB\backslash A}$ of $\sigma_A$ in $\sigma_{AB}$. By definition of the Schur complement, and observing that $\bar{\sigma}_{AB\backslash A}>0$ for any valid CM $\sigma_{AB}$, notice that we can write:  $\sqrt{(\det \sigma_{AB})/(\det \sigma_A)} =\sqrt{\det \bar{\sigma}_{AB\backslash A}} = \prod_{j=1}^{n_B} \bar{\nu}^{AB\backslash A}_j =  {\big(\prod_{j:\bar{\nu}^{AB\backslash A}_j<1} \bar{\nu}^{AB\backslash A}_j\big)\big(\prod_{j:\bar{\nu}^{AB\backslash A}_j \geq 1} \bar{\nu}^{AB\backslash A}_j\big)} \geq {\big(\prod_{j:\bar{\nu}^{AB\backslash A}_j<1} \bar{\nu}^{AB\backslash A}_j\big)}\,.$

Applying $(-\ln)$ to both sides we get, for any CM $\sigma_{AB}$ with ${\cal G}^{A \to B}(\sigma_{AB})>0$, the bound (tight when $n_B=1$ \cite{Kogias:15})
\begin{equation} \label{seqgbomb}
2{\cal G}^{A \to B}(\sigma_{AB}) \geq   {\cal M}(\sigma_A)-{\cal M}(\sigma_{AB})  = -  {\cal I}_{B|A}(\sigma_{AB})\,,
\end{equation}
where ${\cal M}(\sigma) = \ln \det \sigma$ is the log-determinant of the CM $\sigma$ \cite{asimon}, and  ${\cal I}_{B|A}(\sigma_{AB})={\cal M}(\sigma_{AB})-{\cal M}(\sigma_{A})$ is the conditional log-determinant, which  --- in analogy to the standard conditional quantum entropy --- is concave on the set of CMs \cite{asimon} and subadditive with respect to the conditioned subsystems,
\begin{equation}\label{esub}
{\cal I}_{BC|A} (\sigma_{ABC}) \leq {\cal I}_{B|A} (\sigma_{ABC}) + {\cal I}_{C|A} (\sigma_{ABC})\,.
 \end{equation}
 Notice that the latter property is equivalent to the strong subadditivity for the log-determinant of the CM $\sigma_{ABC}$
 \cite{GerardoRenyi,asimon}.

To prove (\ref{monog2}), it suffices to consider the case in which the multimode term ${\cal G}^{A_k \to (A_1,\ldots,A_{k-1},A_{k+1},\dots,A_m)}$ and all the pairwise terms ${\cal G}^{A_k \to A_j}$  are nonzero.
Applying then (\ref{seqgbomb}) to the leftmost term in (\ref{monog2}), and using repeatedly the negation of (\ref{esub}), we get:
${\cal G}^{A_k \to (A_1,\ldots,A_{k-1},A_{k+1},\dots,A_m)}(\sigma_{A_1 \ldots A_m}) \geq \frac12 [{\cal M}(\sigma_{A_1,\ldots,A_{k-1},A_{k+1},\dots,A_m}) - {\cal M}(\sigma_{A_1 \ldots A_m})] = - \frac12 {\cal I}_{(A_1,\ldots,A_{k-1},A_{k+1},\dots,A_m)|A_k} (\sigma_{A_1,\dots,A_m})   \geq  -  \frac12 \sum_{j \neq k}   {\cal I}_{A_j|A_k}(\sigma_{A_1 \ldots A_m}) =  \sum_{j \neq k} {\cal G}^{A_k \to A_j}(\sigma_{A_1 \ldots A_m})$,
where in the last step we used again (\ref{seqgbomb}) which holds with equality on each of the two-mode terms involving $A_k$ and any $A_j$, provided ${\cal G}^{A_k \to A_j}(\sigma_{A_1 \ldots A_m}) > 0$ as per assumption.
This concludes the proof of Eq.~(\ref{monog2}) for any $m$-mode CM  $\sigma_{A_1 \ldots A_m}$. \hfill $\Box$


\begin{thebibliography}{80}
\bibitem [{not({\natexlab{a}})}]{notecleve}%
  \BibitemOpen
  \href@noop {} {} \bibinfo {note} {{Here QSS is intended as
  quantum sharing of a classical secret, not to be confused with sharing a
  quantum state as in \cite{Cleve}}}\BibitemShut {NoStop}%
\end{thebibliography}

%

\end{document}